\documentclass[preprintnumbers,amsmath,amssymb,twocolumn]{revtex4}
\usepackage[colorlinks=false,citecolor=red,filecolor=green,linkcolor=blue,pdfnewwindow=true]{hyperref}
\usepackage{hyperref}
\usepackage{titlesec}
\titleformat{\section}
	{\center\bfseries}{\makebox[30pt][l]{\thesection}}{0pt}{}
\titleformat{\subsection}
	{\flushleft\bfseries}{\makebox[30pt][l]{\thesubsection}}{0pt}{}
\titleformat{\subsubsection}
	{\flushleft\itshape}{\makebox[30pt][l]{\thesubsubsection}}{0pt}{}
\titlespacing\subsection{0pt}{12pt plus 4pt minus 2pt}{0pt plus 2pt minus 2pt}
\titlespacing\subsubsection{0pt}{12pt plus 4pt minus 2pt}{0pt plus 2pt minus 2pt}
\usepackage[font=small]{caption}
\usepackage{subcaption}
\usepackage{color}
\usepackage{graphicx}
\usepackage{makeidx}
\usepackage{bm}
\usepackage[title]{appendix} 
\usepackage{tabu}
\usepackage{float}
\usepackage{chemarr}
\usepackage{url}

\begin{document}

\title{Pre-oligomerisation stochastic dynamics of prions driven by water molecules}
\author{Mairembam Kelvin Singh$^{1}$}
\email{kelvin.phd.phy@manipuruniv.ac.in}
\author{R.K. Brojen Singh$^2$}
\email{brojen@jnu.ac.in}
\author{Moirangthem Shubhakanta Singh$^1$}
\email{mshubhakanta@gmail.com}
\affiliation{$^1$Department of Physics, Manipur University, Canchipur-795003, Imphal, Manipur, India.\\ $^2$School of Computational $\&$ Integrative Sciences, Jawaharlal Nehru University, New Delhi-110067, India.}

\begin{abstract}
{\noindent}Prions are proteinaceous infectious particles that cause neurodegenerative diseases in humans and animals. The complex nature of prions, with respect to their conformations and aggregations, has been an important area of research for quite some time. Here, we develop a model of prion dynamics prior to the formation of oligomers and subsequent development of prion diseases within a stochastic framework, based on the analytical Master Equation and Gillespie's Stochastic Simulation Algorithm (SSA). The results that we obtain shows that solvent water molecules act as driving agents in the dynamics of prion aggregation. Further, it is found that aggregated and non-aggregated proteins tend to co-exist in an equilibrium state, depending upon the reaction rate constants. These results may provide a theoretical and qualitative contexts of possible therapeutic strategies in the treatment of prion diseases.\\ 

\noindent\textbf{Keywords:} Prions; Protein aggregation; Master Equation; Stochastic Simulation. 
\end{abstract}
\maketitle

\section{Introduction}
{\noindent}Prions are proteinaceous infectious particles that cause neurodegenerative diseases in humans and animals. Notable diseases caused by prions are the Bovine spongiform encephalopathy (BSE), scrapie of sheep and Creutzfeldt-Jakob disease (CJD) of humans. Prions lack nucleic acid and the prion diseases modify the normal cellular prion protein abbreviated as PrP$^{\text{C}}$ into the modified form abbreviated as PrP$^{\text{Sc}}$ \cite{Prusiner}. This conversion takes place when they have similar amino acid sequences \cite{Prusiner, Cohen}. PrP$^{\text{C}}$ contains mainly $\alpha$-helix isoform and is protease-sensitive while PrP$^{\text{Sc}}$ is mainly composed of $\beta$-sheet isoform and is protease-resistant \cite{Prusiner, Srivastava, Horwich}.\\

{\noindent}The spread of prion diseases from one species to another is a stochastic process, where the dynamics of the prions population is governed by probabilistic equation(s). This process also features the formation of intermediates \cite{Cohen} and a long incubation period from the infected time to the onset of symptoms \cite{Prusiner}. Studies have also stated that the diseases are caused by protein misfolding and aggregation, when monomeric protein chains diffuse within the same molecule \cite{Srivastava, Lapidus}. A unique characteristic of prion diseases is that the aggregated prions promote further aggregation of normal proteins, thus making it possible for transmission of the diseases \cite{Horwich}.\\

{\noindent}Therapeutic control strategies like using a particular molecule to stabilise the structure of PrP$^{\text{C}}$ or modifying the action of the intermediates formed during the system reactions, were already suggested \cite{Prusiner}. These molecules must be able to prevent aggregation to control the transmission of diseases \cite{Lapidus}. However, one of the difficulties that arises during the control strategies is that prions occur in different ``strains", producing different patterns and complex behaviors \cite{Horwich, Lemarre}. The lack of nucleic acid also makes the prions harder to remove by ultraviolet irradiation \cite{Mobley}. So, studying the complex behavior of prion aggregation is crucial for developing a sustainable control strategy \cite{Lapidus, Chong}.\\

{\noindent}When an ensemble of monomeric unfolded prion conformations is subjected to a large number of solvent water molecules, oligomerisation or the formation of oligomers takes place. Two aggregated prions can also form a dimerised protein (dimer) through molecular diffusion, followed by the production of oligomers, which may be toxic to cells \cite{Lapidus, Stephens, Chong}. These oligomers, then, eventually form amyloid fibrils \cite{Chong, Eisenberg}. The water molecules play a big role in controlling the dynamics of the prions. The hydration shell present in prion proteins controls the movement and function of the proteins and the removal of which can drive protein folding \cite{Stephens, Verma, Jamal, Uversky}.\\ 

{\noindent}Various mathematical models and computational simulations to study the aggregation of prions have been done in recent years. In \cite{Matthaeus}, brain infection by prion diseases has been studied by developing a mathematical model based on the reaction-diffusion equations considering the interaction of prions with transport on networks. Qualitative studies have shown the ability of the prions to co-exist in a stable state influenced by initial conditions applied to the mathematical model developed \cite{Lemarre}. From a stochastic point of view, a stochastic cellular automata model has been developed in \cite{Mobley} based on the probabilistic nature of interaction and diffusion. Liu and Kang \cite{Liu} studied the protein aggregation by developing the chemical master equation and applying the lognormal moment closure method with the Stochastic Simulation Algorithm (SSA) by Gillespie \cite{Gillespie} confirming its accuracy.\\

{\noindent}In this paper, we also studied the prion aggregation dynamics using the Master Equation approach and solve it by the generating function technique \cite{McQuarrie, Gardiner} to obtain the Poissonian probability distribution of dimers, formed during the process of aggregation and development of oligomers. We also used Gillespie's SSA \cite{Gillespie} to run various simulations based on different reaction rates and initial conditions of the solvent water molecules and non-aggregated prion proteins.\\

{\noindent}We found that the asymptotic behavior of dimers follow a Poisson process and the mean population of dimers also saturate to a value equal to the ratio of the mean production rate to the mean degradation rate. This saturation is also confirmed by the SSA simulation. Further, we found that certain reaction rates control the co-stability and co-existence \cite{Lemarre} of aggregated and non-aggregated prions, without producing significant amount of oligomers in the process.\\

{\noindent}The paper is organised as follows. Section II describes the model reactions depicting the various processes involved during the aggregation. Section III briefly describes the theoretical context of the analytical and computational methods used. Section IV shows the various analytical and computational simulation results that we obtain and its subsequent analysis for the significance in providing possible qualitative control strategies.\\

\section{Model Description}
{\noindent}We consider an ensemble of well-mixed unfolded prion monomers subjected to a large number of solvent water molecules $(H)$. Let $X$ denote the non-aggregated prions and $X^*$ denote the aggregated prions. If two $X^*$ interact through molecular diffusion, a dimer $D$ is formed. This is followed by the formation of an oligomer $O$ or production of $X$ and $X^*$. These various processes \cite{Srivastava, Stephens} are then represented in the form of a system of reactions as shown below:
\begin{eqnarray}
\label{re}
\begin{split}
X+H&\stackrel{\rm k_1}{\longrightarrow} X^* + H\\
X^* + H&\stackrel{\rm k_2}{\longrightarrow} X+H\\
2X^* &\stackrel{\rm k_3}{\longrightarrow} D\\
D&\stackrel{\rm k_4}{\longrightarrow} X+X^* \\
D&\stackrel{\rm k_5}{\longrightarrow} O+H\\
O&\stackrel{\rm k_6}{\longrightarrow} D
\end{split}
\end{eqnarray}
{\noindent}The first two reactions denote the inter-conversions of aggregated $(X^*)$ and non-aggregated $(X)$ prions through the help of the water molecules $(H)$ with reaction rate constants $k_1$ and $k_2$. The third reaction denotes the formation of a dimer $(D)$ from two aggregated prions $(X^*)$ with reaction rate constant $k_3$. The fourth reaction denotes the conversion of a dimer $(D)$ into an aggregated $(X^*)$ and non-aggregated $(X)$ prion with reaction rate constant $k_4$. The fifth reaction denotes the formation of an oligomer $(O)$ and water $(H)$ from a dimer $(D)$ with reaction rate constant $k_5$. The last reaction denotes the conversion of an oligomer $(O)$ to a dimer $(D)$ with reaction rate constant $k_6$.\\

{\noindent}Throughout our study, we assume that the system is a well-mixed and well-stirred system, with the volume taken as $V=1$ so that the stochastic and deterministic rate constants are equal.\\

\section{Methods}
{\noindent}The system of reactions in \eqref{re} is studied and analysed through two approaches, the analytical master equation approach \cite{McQuarrie, Gardiner} and the computational simulation using Gillespie's SSA \cite{Gillespie}. These approaches are briefly described in the next subsections.\\

\subsection{Analytical Method}
{\noindent}The dynamics of a particle in a stochastic system can be described by the Chapman-Kolmogorov (CK) equation \cite{Kolmogorov, Gardiner} where the state transitions obey Markov process. The CK equation in integral form for a stochastic particle transitioning from a state $(y^\prime,t^\prime)$ to a state $(y,t)$ is given by,
\begin{eqnarray}
\label{cke}
\frac{\partial}{\partial t}P(y,t|y^\prime,t^\prime)&=&-\sum_i \frac{\partial}{\partial u_i}\left[A_i(u,t)P(u,t|y^\prime,t^\prime)\right]\nonumber\\
&&+\frac{1}{2}\sum_{i,j}\frac{\partial^2}{\partial u_i\partial u_j}\left[B_{ij}(u,t)P(u,t|y^\prime,t^\prime)\right]\nonumber\\
&&+\int dy [W(u|y,t)P(y,t|y^\prime,t^\prime)\nonumber\\
&&-W(y|u,t)P(u,t|y^\prime,t^\prime)]
\end{eqnarray}
where $\left\lbrace W\right\rbrace$ are Wiener functions of transition probabilities of the states, $A_i$ and $B_{ij}$ are the functions depending on the properties of the system and $P(y,t|y^\prime,t^\prime)$ is the jump probability. The master equation , which is the discrete form of the CK equation, derived from the integral part of the CK equation \eqref{cke}, is given by \cite{Kolmogorov, Gardiner, Kampen},
\begin{eqnarray}
\label{me}
\frac{\partial}{\partial t}P(y,t|y^\prime,t^\prime)&=&\int dy [W(u|y,t)P(y,t|y^\prime,t^\prime)\nonumber\\
&&-W(y|u,t)P(u,t|y^\prime,t^\prime)]
\end{eqnarray}

{\noindent}The master equation can be considered as the time evolution of the probability of a particle in a system obeying Markov process \cite{Singh}. We can solve this master equation \eqref{me} by the generating function (GF) technique \cite{McQuarrie}. The $n$-dimensional GF of the probability distribution function $P(y_1,y_2,...;t)$ is given by,
\begin{eqnarray}
\label{gf}
G(s_1,s_2,...,s_n)=\sum_{y_1,y_2,...,y_n}\prod_{i=1}^n s_i^{y_i}P(y_1,y_2,...,y_n;t)
\end{eqnarray}

{\noindent}Multiplying the master equation \eqref{me} by $\displaystyle \sum_{y_1,y_2,...,y_n}\prod_{i=1}^n s_i^{y_i}$ and using \eqref{gf}, we can transform the master equation \eqref{me} to a spatio-temporal PDE of the GF with the boundary condition, $\displaystyle G(s_1,s_2,...,s_n;0)=\prod_{i=1}^n s_i^{N_i^{[0]}}$, where $\{N_i^{[0]}\};i=1,2,...,n$ is the set of initial populations of the respective variables $y_1,y_2,...y_n$. After putting back the solution of GF to \eqref{gf} and equating the coefficients, we can obtain the solution $P(y_1,y_2,...,y_n;t)$ of the master equation \eqref{me} \cite{Singh}.\\

\subsection{Computational Simulation}
{\noindent}For computational simulation, we use the stochastic simulation algorithm (SSA) by Gillespie \cite{Gillespie}, which is based on the assumption that the time evolution of a stochastic particle is a discrete process obeying Markov property. The population of the particle increases or decreases by integral amounts according to the reactions occurring in the system.\\

{\noindent}Consider a system of $n$ molecular species reacting in $m$ different ways. The joint reaction probability density function for a random reaction $r$ and reaction time $T$ \cite{Gillespie} is then given as,
\begin{eqnarray}
\label{PDF}
P(r,T)= \left\{
  \begin{array}{ll} 
      a_{r} \exp(-a_0T), & ~ \text{if} ~ 0\le T < \infty \\
      & \text{and} ~ r = 1,...,m \\
      0, & \text{otherwise} 
      \end{array}
\right.
\end{eqnarray}
The propensity function of the $r$th reaction is given by,
\begin{eqnarray}
\label{a}
a_{r} = h_{r}k_{r} ~~\forall~ r\in m;~~~~
a_0 =\sum_{r=1}^m a_{r}=\sum_{r=1}^m h_{r}k_{r}
\end{eqnarray}
As $P(r)$ and $P(T)$ are independent of each other, we can rewrite \eqref{PDF} as \cite{Gillespie},
\begin{eqnarray}
\label{P1P2}
\begin{split}
P(r,T) &= P(r) P(T) \\
\text{where}~~P(T) &= a_0 \exp(-a_0 T) \\
\text{and}~~P(r) &= \frac{a_{r}}{a_0}
\end{split}
\end{eqnarray}
From \eqref{P1P2}, we can generate two uniform random numbers $R_1$ and $R_2$ using a unit interval uniform random number generator for $P(r)$ and $P(T)$ respectively for simulation,
\begin{eqnarray}
\label{tau}
\begin{split}
T &= \left(\frac{1}{a_0}\right) \ln \left(\frac{1}{R_1}\right)\\
\sum_{\nu=1}^{r-1} a_{\nu} &< R_2 a_0 \le \sum_{\nu=1}^{r} a_{\nu}
\end{split}
\end{eqnarray}
The algorithm of the SSA is as follows:
\begin{enumerate}
\item Initialise the values of the $m$ reaction constants $k_1,...,k_m$ and the $n$ initial population of the species involved. The time variable $t$ and the reaction counter $n'$ are both set to zero and the unit interval uniform random number generator is also imported.
\item Calculate the $m$ propensities $a_1=h_1k_1,...,a_m=h_mk_m$ and $a_0$ using \eqref{a}.
\item According to the unit interval uniform random number generator imported earlier, two random numbers $R_1$ and $R_2$ are generated, $r$ and $T$ are calculated, using \eqref{tau}.
\item Increase $t$ by $T$, and change the population of the species according to the particular reaction.
\item Go back to step 2, recalculate $a_{\nu}$, $a_0$ and then proceed.
\item Stop the loop when either $t$ or $n'$ reaches the desired value, or if $a_0$ becomes zero.\\
\end{enumerate}

\begin{figure*}
\label{Fig1}
\begin{center}
\includegraphics[height=12.0cm,width=18.0cm]{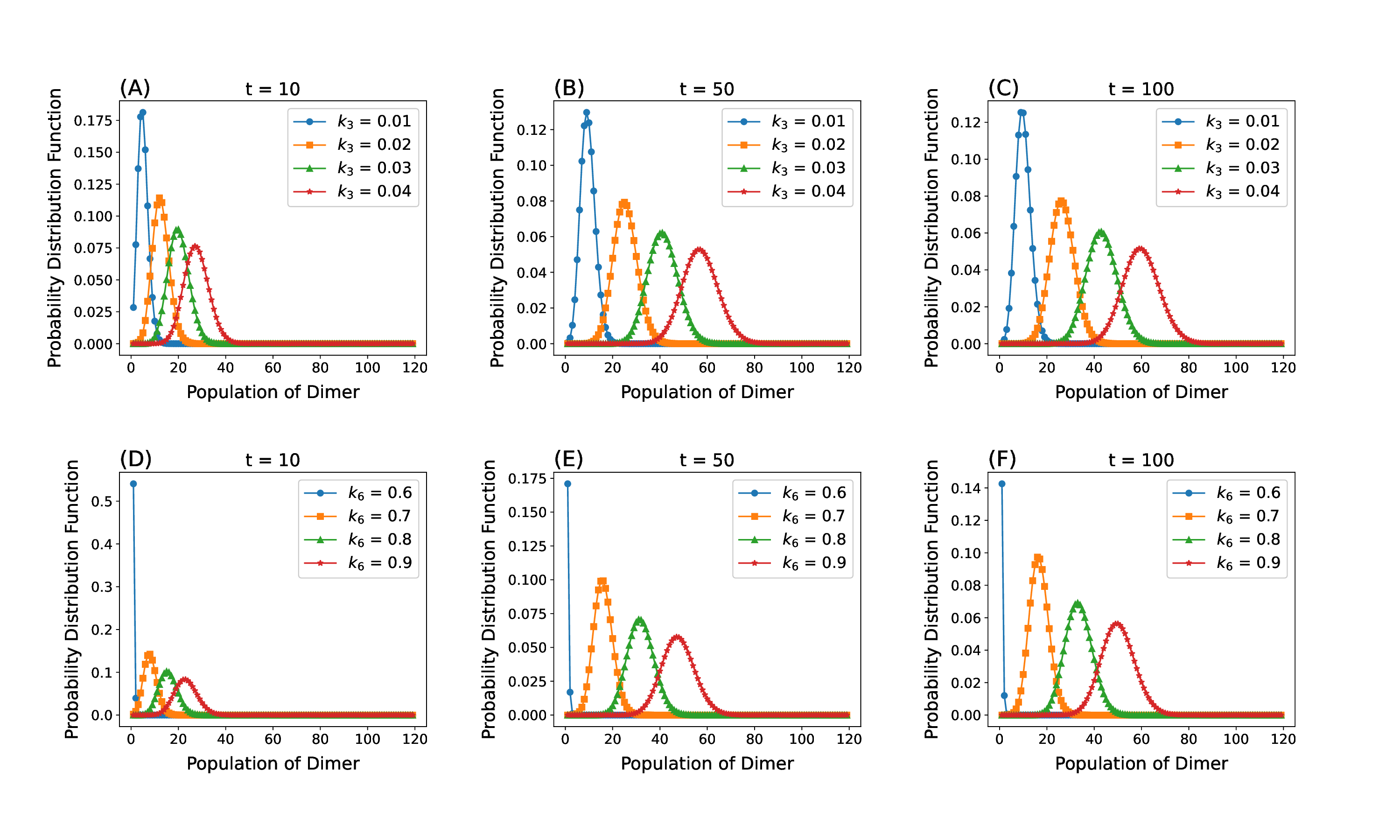}
\caption{Plots of the PDF against the dimer population for different values of time $t=10,50,100$ units. \textbf{(A)-(C)}: These are plots corresponding to the variation of $k_3$. As $k_3$ increases, the peak value decreases and the distribution shifts towards the larger dimer population. Also, as time increases, the distribution moves towards the larger dimer population but saturates to a fixed pattern and the peak value corresponding to each $k_3$ decreases. \textbf{(D)-(F)}: These are plots corresponding to the variation of $k_6$. We can see a similar behavior as in the case of variation of $k_3$.}
\end{center}
\end{figure*}

\begin{figure*}
\label{Fig2}
\begin{center}
\includegraphics[height=10.0cm,width=16.0cm]{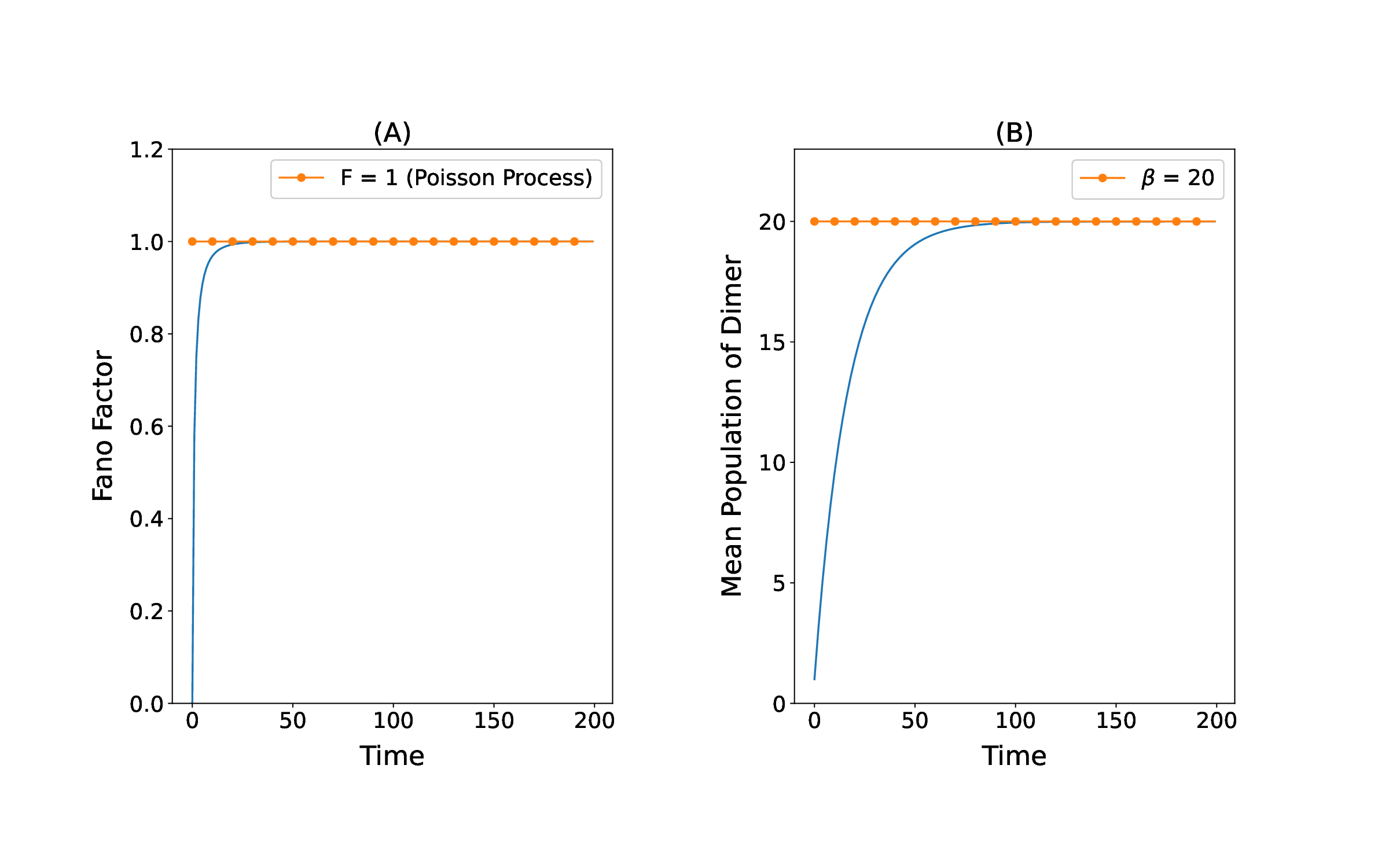}
\caption{\textbf{(A)} Fano factor of prion aggregation dynamics as a function of time. It saturates to the value 1, implying that the dynamics is a Poisson process. \textbf{(B)} Mean population of dimer as a function of time. The mean population saturates to a value equal to $\beta$.}
\end{center}
\end{figure*}

\section{Results and Discussion}
{\noindent}Based on the two methods used, the results are divided into two parts, an analytical one and a simulation result. The analytical result gives the qualitative analysis of the aggregation dynamics in terms of the probability distribution function of the dimers $(D)$ formed during the prion aggregation. On the other hand, the simulation study considers other species such as the aggregated prions $(X^*)$, non-aggregated prions $(X)$ and oligomers $(O)$ apart from the dimers $(D)$. So, the simulation study gives a more broader picture of the time evolution of the different molecular species considered in the system.\\

\subsection{Analytical Results}
{\noindent}To derive the master equation of the dimers $(D)$ and study their dynamics, we can ignore the first two reactions of \eqref{re} as they do not involve dimers $(D)$. Considering the state transitions based on the system of reactions in \eqref{re} and using the principle of detailed balance, the probability distribution function of the dimer population $(D)$ at a particular time $t+\Delta t$, provided that the previous state occurred in time $t$, is given by \cite{Singh},
\begin{eqnarray}
P(D;t+\Delta t)&=&\frac{X^*\left(X^*-1\right)}{2}k_3 P(D-1;t)\Delta t\nonumber\\
&&+\left(D+1\right)k_4 P(D+1;t)\Delta t\nonumber\\
&&+\left(D+1\right)k_5 P(D+1;t)\Delta t\nonumber\\
&&+Ok_6 P(D-1;t)\Delta t\nonumber\\
&&+\left[1-\omega\Delta t\right]P(D;t)\nonumber
\end{eqnarray}
where $\displaystyle\omega=\frac{X^*\left(X^*-1\right)}{2}k_3+Dk_4+Dk_5+Ok_6$.

{\noindent}Applying the limit as $\Delta t \rightarrow 0$, we can obtain the master equation as,
\begin{eqnarray}
\label{meD}
\frac{\partial P(D;t)}{\partial t}&=&\frac{X^*\left(X^*-1\right)}{2}k_3 P(D-1;t)\nonumber\\
&&+\left(D+1\right)k_4 P(D+1;t)\nonumber\\
&&+\left(D+1\right)k_5 P(D+1;t)\nonumber\\
&&+Ok_6 P(D-1;t)\nonumber\\
&&-\omega P(D;t)
\end{eqnarray}

{\noindent}Now, using the generating function, $\displaystyle G(s,t)=\sum_D s^D P(D;t)$, with the initial condition, $\displaystyle G(s,0)=s^M$, where $M$ is the initial dimer population, we obtain the solution of the master equation \eqref{meD} as,
\begin{eqnarray}
\label{PDFD}
P(D;t)&=&e^{\beta\left(e^{-\alpha t}-1\right)}\sum_i ^M C_i \frac{1}{(D-i)!}\nonumber\\
&&e^{-i\alpha t}\beta^{D-i}\left(1-e^{-\alpha t}\right)^{M+D-2i}
\end{eqnarray}
where $\alpha=k_4+k_5$ and $\displaystyle \beta=\frac{X^*(X^*-1)k_3+2Ok_6}{2\alpha}$, the ratio of the mean production rate to the degradation rate of dimers.\\

{\noindent}The probability distribution function (PDF) obtained in \eqref{PDFD} is plotted with respect to the dimer population in Figure 1. Upon increase of dimer production from aggregated prions, i.e., increase of $k_3$, the peak value decreases and the distribution shifts towards the larger dimer population. Also, as time increases, the distribution moves towards the larger dimer population but saturates to a fixed pattern, which can be seen by the close similarities of Figures 1(B) and 1(C). The peak value corresponding to each $k_3$ also decreases as time increases. We can also see the same behavior if we increase the conversion of oligomer into dimer, i.e., increase $k_6$.\\

{\noindent}To analyse the distribution pattern, we can simplify the expression \eqref{PDFD} with the approximation $M-i\approx M$ and $D-i\approx D$ as,
\begin{eqnarray}
\label{Pois}
P(D;t)=(1-\gamma)^M e^\gamma \textit{Pois}(\delta)
\end{eqnarray}
where $\displaystyle\gamma=e^{-\alpha t}$, $\delta=\beta(1-\gamma)$ and \textit{Pois}$(\delta)$ denotes the Poisson distribution function in $\delta$. This deduction proves that the prion aggregation dynamics is a Poisson process.\\

\subsubsection*{Observables -}
{\noindent}To study the influence of noise in the prion aggregation dynamics, we calculated the Fano factor \cite{Fano}, using the relation, $\displaystyle \mathcal{F}=\frac{\sigma^2}{\langle D \rangle}$, where $\sigma^2$ is the standard deviation and $\langle D \rangle$ is the mean population of dimers.\\

{\noindent}The mean of the dimer population is determined from the GF using the relation $\displaystyle \langle D \rangle = \left.\frac{\partial G(s,t)}{\partial s}\right|_{s=1}$. It is obtained as,
\begin{eqnarray}
\label{meanD}
\langle D\rangle=(M-\beta)\gamma +\beta
\end{eqnarray}

{\noindent}The asymptotic values of the mean dimer population can be easily determined as, $\displaystyle \lim_{t\rightarrow 0}\langle D\rangle=M$, the initial dimer population and $\displaystyle \lim_{t\rightarrow \infty}\langle D\rangle=\beta$, the ratio of the mean production rate to the degradation rate of dimers.\\

{\noindent}The standard deviation $\sigma^2$ can be determined using the relation,
\begin{eqnarray}
\label{SD}
\sigma^2=\langle D^2 \rangle -\langle D\rangle^2
\end{eqnarray}
where $\displaystyle \langle D^2 \rangle=\left.\frac{\partial^2 G(s,t)}{\partial s^2}\right|_{s=1}+\langle D\rangle$. From \eqref{meanD} and \eqref{SD}, the Fano factor is determined as,
\begin{eqnarray}
\label{Fano}
\mathcal{F}=\frac{\left(M\gamma +\beta\right)\left(1-\gamma\right)}{\left(M-\beta\right)\gamma +\beta}
\end{eqnarray}

{\noindent}The asymptotic values of the Fano factor is also determined as, $\displaystyle \lim_{t\rightarrow 0}\mathcal{F}=0$ and $\displaystyle \lim_{t\rightarrow \infty}\mathcal{F}=1$. These values also show that the prion aggregation dynamics is a Poisson process \cite{Rajdl, Chanu}. Thus, there is no specific role of noise in the prion aggregation dynamics. The asymptotic behavior of the Fano factor and the mean dimer population are also shown in Figure 2.\\

\begin{figure*}
\label{Fig3}
\begin{center}
\includegraphics[height=12.0cm,width=18.0cm]{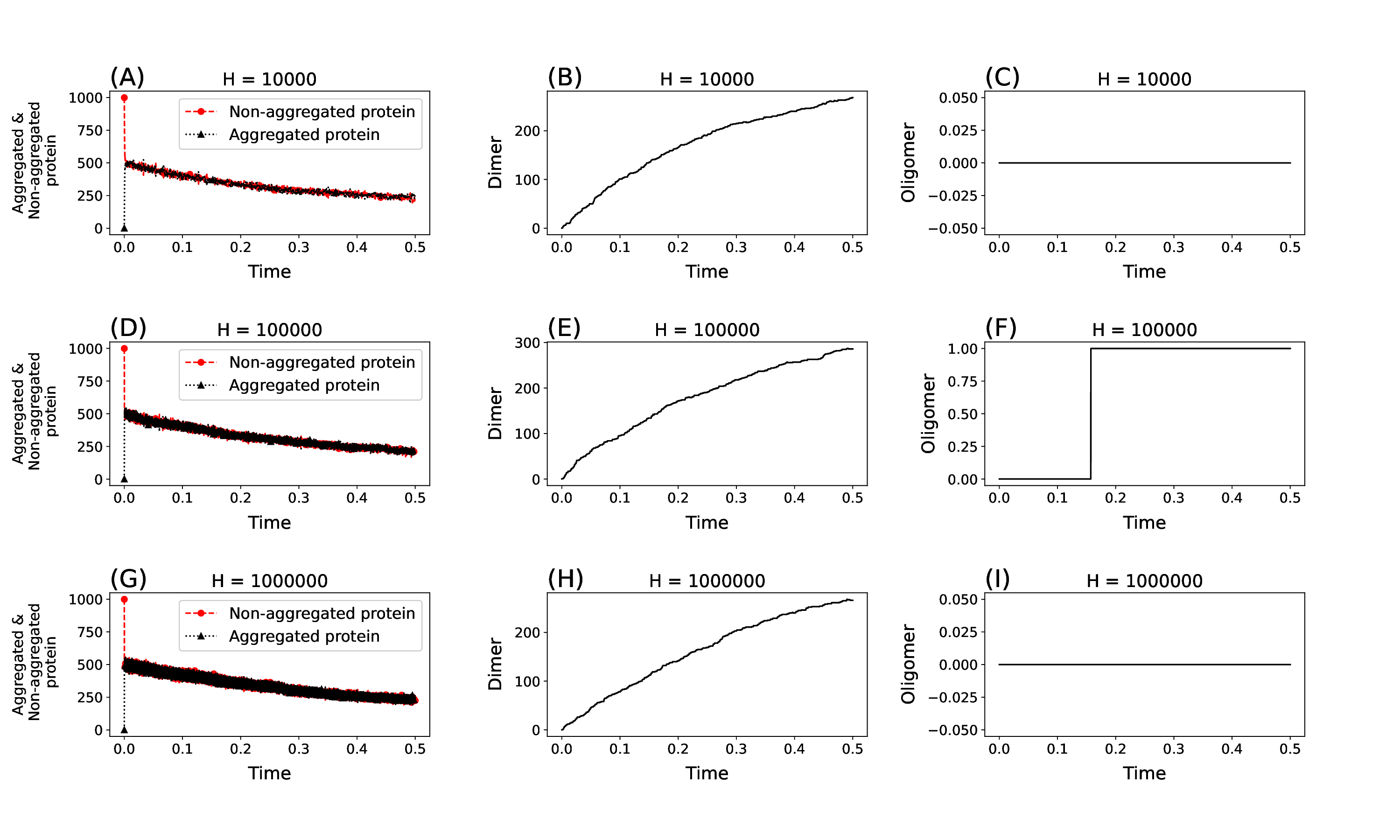}
\caption{The aggregated and non-aggregated prions approach towards a stable value at a very quick time and their populations cluster together showing signs of co-existence towards an equilibrium state (left column). The dynamics of dimer population remain similar to the analytical result and the final population also approaches towards the common stable value of the aggregated and non-aggregated prions (middle column). There is very little to almost no oligomerisation (right column).}
\end{center}
\end{figure*}

\subsection{Simulation Results}
{\noindent}Simulations using SSA are done to the reaction system \eqref{re} in different ways to analyse the aggregation dynamics from different perspectives. For the first simulation, the initial population of water molecules are varied. The values of the reaction rate constants used are: $k_1=0.1$, $k_2=0.1$, $k_3=0.01$, $k_4=0.05$, $k_5=0.01$ and $k_6=0.6$. The initial populations of the different species involved are set as, water, $H=(10^4,10^5,10^6)$; non-aggregated proteins, $X_{in}=1000$; aggregated proteins, $X^*_{in} =0$; dimer, $D_{in}=0$ and oligomer, $O_{in}=0$. In this simulation, we kept $k_1>k_3$, i.e., the rate of formation of aggregated prion proteins is more than the formation of dimers. The aggregated and non-aggregated prions approach towards a stable value at a very quick time and their populations cluster together showing signs of co-existence towards an equilibrium state. The dynamics of dimer population remain similar to what we obtained in the analytical result. The final population of dimers also approaches towards the common stable value of the aggregated and non-aggregated prions. There is little to almost no oligomerisation. As $H$ increases by a factor of 10, the total number of reactions for the fixed amount of time increases by a factor of 10. The total number of reactions for the respective initial water populations are of the order of $(10^5,10^6,10^7)$. This means that the population of water molecules determines the total reaction rate of the system. Most of the reactions are mainly the first two reactions, which also signify that water controls oligomerisation process during the aggregation. These results are shown in Figure 3.\\

\begin{figure*}
\label{Fig4}
\begin{center}
\includegraphics[height=8.0cm,width=18.0cm]{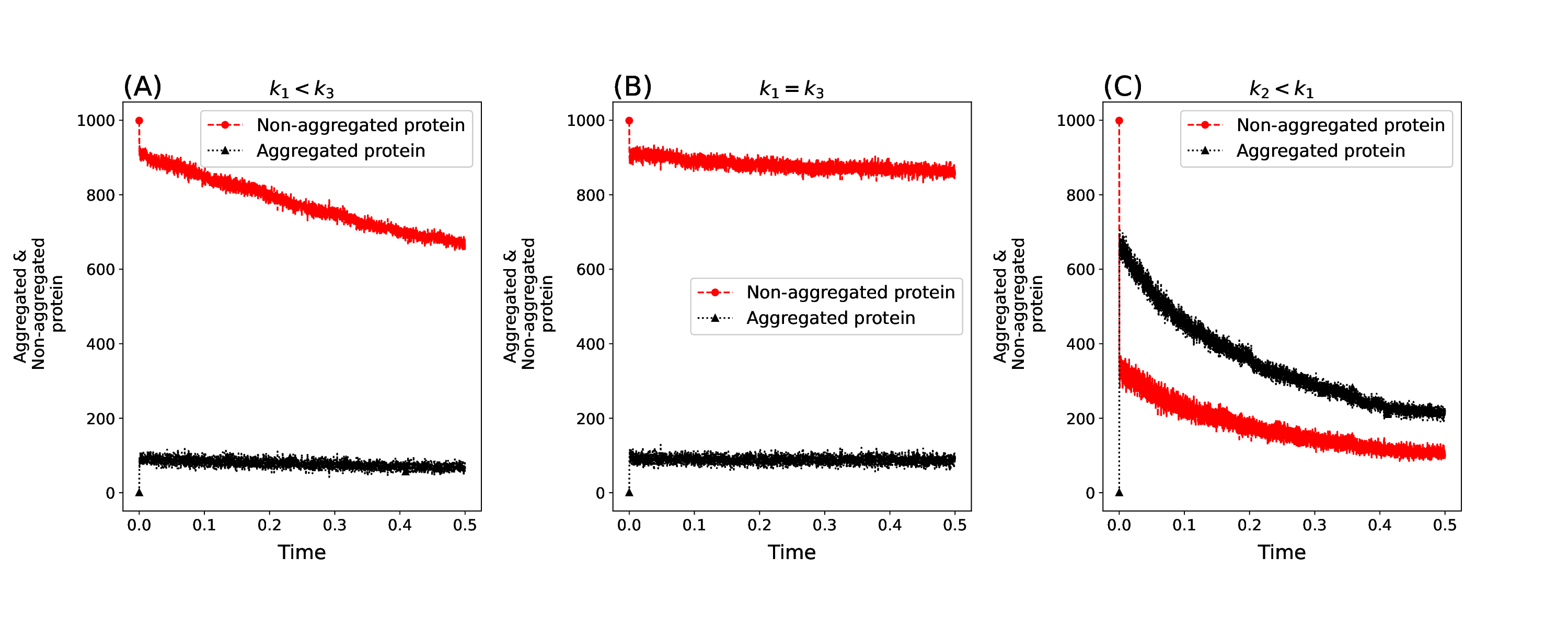}
\caption{\textbf{(A) and (B):} The populations of aggregated and non-aggregated proteins no longer approach towards a common stable state. The population of aggregated prions oscillate around their stable state, the non-aggregated prions seem to decline towards the direction of aggregated prions indicating the possibility of aggregated and non-aggregated prions approaching co-stability at a much later time. \textbf{(C):} The population of aggregated prions exceeds that of the descending population of non-aggregated prions but eventually decreases to achieve the co-stability, sooner than the first two cases.}
\end{center}
\end{figure*}

{\noindent}Next, we keep the population of water molecules fixed at $H=10^5$ and observe the aggregation dynamics at different variations of the rate constants at the same range of time. For $k_1<k_3$ and $k_1=k_3$, the populations of aggregated and non-aggregated prion proteins no longer approach towards a common stable state (Figure 4A and 4B). However, the distinguishing feature is that although the population of aggregated prions oscillate around their stable state, the non-aggregated prions seem to decline towards the direction of aggregated prions. This indicates the possibility of aggregated and non-aggregated prions approach co-stability at a much later time. For the case of $k_2<k_1$, i.e., the formation of aggregated prion proteins from the non-aggregated prion proteins occurring at a faster rate than the reverse process, the population of aggregated prions exceeds that of the descending population of non-aggregated prions but eventually decreases to achieve the co-stability, sooner than the first two cases (Figure 4C). In all three cases, the total number of reactions is of the order of $10^5$.\\

\begin{figure*}
\label{Fig5}
\begin{center}
\includegraphics[height=12.0cm,width=18.0cm]{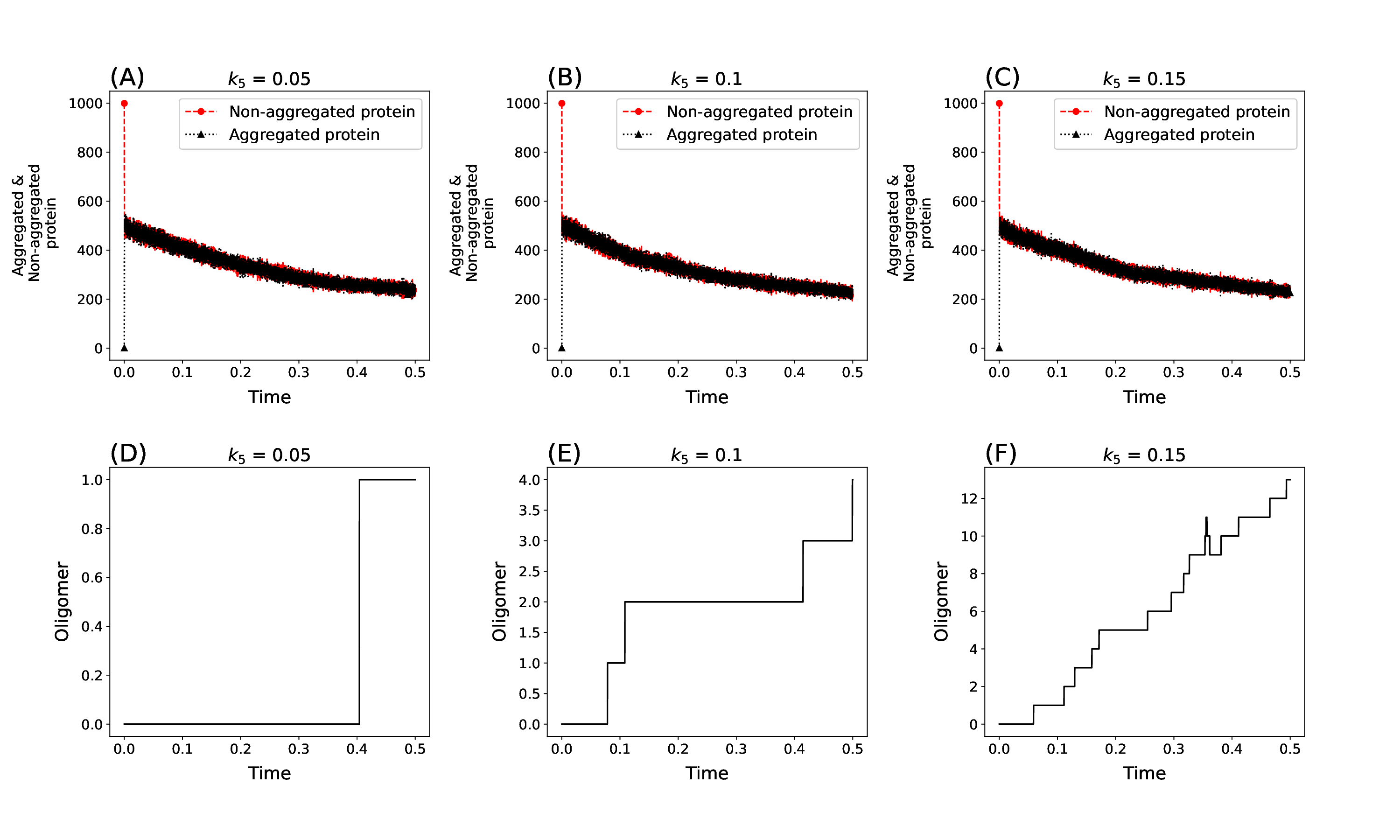}
\caption{\textbf{(A)-(C):} The aggregated and non-aggregated prions approach together towards a stable value and their populations cluster together showing signs of co-existence towards an equilibrium state. \textbf{(D)-(F):} Oligomer formation starts and increases as $k_5$ increases.}
\end{center}
\end{figure*}

{\noindent}Now, the production rate of oligomers $(k_5)$ is increased with $k_1>k_3$ and $H=10^5$. The dynamics of aggregated and non-aggregated prion proteins remain similar and as expected, oligomer formation starts and increases as $k_5$ increases (Figure 5). The total number of reaction is of the order of $10^6$.\\

\begin{figure*}
\label{Fig6}
\begin{center}
\includegraphics[height=12.0cm,width=18.0cm]{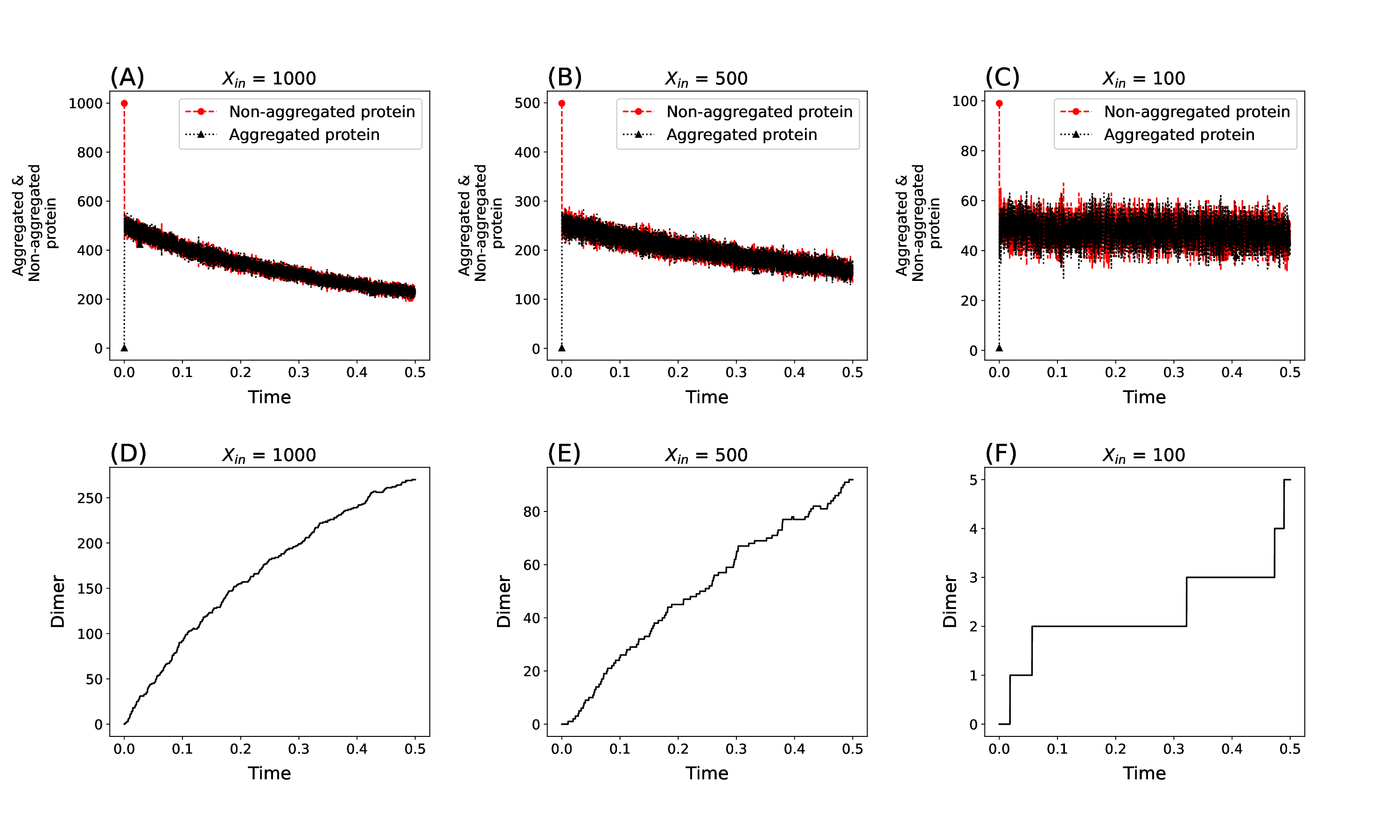}
\caption{\textbf{(A)-(C):} The aggregated and non-aggregated prions approach together towards a stable value and their populations cluster together showing signs of co-existence towards an equilibrium state. \textbf{(D)-(F):} The population of dimers saturate at a quicker time as the initial non-aggregated protein population decreases. For $X_{in}=100$, the dimer population increases discretely and there are instances where the population remains constant for a certain period of time.}
\end{center}
\end{figure*}

{\noindent}Finally, the initial population of non-aggregated proteins is decreased from 1000 to 500 and then to 100. The aggregated and non-aggregated prion dynamics remain similar again. However, the dynamics of dimers begin to change. They saturate at a quicker time as the initial non-aggregated protein population decreases. Particularly, for $X_{in}=100$, the dimer population increases in discrete steps and there are instances where the population remains constant for a certain period of time (Figure 6). The total number of reactions also decreases from the order of $10^6$ to $10^5$ as $X_{in}$ decreases. This means that, like the water molecules, the initial population of non-aggregated proteins controls the total reaction rate of the system.\\

\section{Conclusion}
{\noindent}From the study of the aggregation dynamics of prion proteins, we are able to understand some qualitative behavior of the aggregation before oligomerisation sets in. The stochastic analysis shows the sensitive dependence of aggregation on the initial populations of the involved species. Specifically, we have found that the water molecules and the initial population of non-aggregated prions control the rate of total reactions of the system. The process of oligomerisation only starts when the rate of oligomer formation is increased. Inhibiting the rate of oligomerisation is a way to control the consequential stages of disease transmission. In all the different observations made, the presence of water controls the entire reaction system. This is evident from the fact that out of all the total number of reactions, huge number comes from the inter-conversion of aggregated to non-aggregated prion proteins through water molecules and vice-versa. The rate of certain reactions also determine the state of co-stability and co-existence between the aggregated and non-aggregated prion proteins.\\

{\noindent}The analytical method is limited to the dynamics of dimers only. However, it produces an important result that the dimer dynamics, and hence the aggregation dynamics is a Poisson process, where noise or fluctuations have no significant impact. The asymptotic results showing Poissonian dynamics also describes the random nature of the dynamics.\\

{\noindent}On the other hand, the computational simulations by SSA enable us to understand the aggregation dynamics, by considering the important parameters like the reaction rates and the initial conditions. However, few features like time delay and larger system size are neglected to simplify the study. Although the aggregation occurs at a very quick time of the order of nanoseconds, there may be certain time lag between the reactions. For this, we need to use the delay algorithms \cite{Bratsun, Barrio, Cai, Anderson} to make realistic visualisations. Moreover, we are able to show the qualitative picture of how the aggregation dynamics behave according to changing parameters.\\

{\noindent}{\bf Acknowledgements} \\
{\noindent}MKS is a Junior Research Fellow (JRF) under the National Fellowship for Scheduled Castes Students (NFSC) scheme implemented by the National Scheduled Castes Finance and Development Corporation (NSFDC) under the Department of Social Justice \& Empowerment, Ministry of Social Justice \& Empowerment, Government of India with UGC ref. no. 201610025460. The simulations were done through Python 3 programs written using the Spyder environment of Anaconda distribution. Simulations that require large computational tasks were done in HPE ProLiant DL380 Gen10 Plus installed in the Department of Physics, Manipur University.\\

{\noindent}{\bf Author Contributions:}\\
{\noindent}RKBS and MSS conceptualized the work. MKS did the analytical, computational work and preparation of the figures. MKS, MSS and RKBS wrote the manuscript, read, analyzed the results, and approved the final manuscript.\\

{\noindent}{\bf Additional Information} \\
\textbf{Competing interests:} \\ The authors declare no competing interests.


\begin{thebibliography}{99}
\bibitem{Prusiner}Prusiner, S.B. (1998) Prions. Proc. Natl. Acad. Sci. 95(23), 13363-13383.
\bibitem{Cohen}Cohen, F.E., Pan, K.M., Huang, Z., Baldwin, M., Fletterick, R.J. \& Prusiner, S.B. (1994) Structural clues to prion replication. Science, 264(5158), 530-531.
\bibitem{Srivastava}Srivastava, K.R. \& Lapidus, L.J. (2017) Prion protein dynamics before aggregation. Proc. Natl. Acad. Sci., 114(14), 3572-3577.
\bibitem{Horwich}Horwich, A.L. \& Weissman, J.S. (1997) Deadly conformations—protein misfolding in prion disease. Cell, 89(4), 499-510.
\bibitem{Lapidus}Lapidus, L.J. (2013) Understanding protein aggregation from the view of monomer dynamics. Mol. BioSyst., 9(1), 29-35.
\bibitem{Lemarre}Lemarre, P., Pujo-Menjouet, L. \& Sindi, S.S. (2019) Generalizing a mathematical model of prion aggregation allows strain coexistence and co-stability by including a novel misfolded species. J. Math. Biol., 78, 465-495.
\bibitem{Mobley}Mobley, D.L., Cox, D.L., Singh, R.R., Kulkarni, R.V. \& Slepoy, A. (2003) Simulations of Oligomeric Intermediates in Prion Diseases. Biophys. J., 85(4), 2213-2223.
\bibitem{Chong}Chong, S.H. \& Ham, S. (2015) Distinct role of hydration water in protein misfolding and aggregation revealed by fluctuating thermodynamics analysis. Acc. Chem. Res., 48(4), 956-965.
\bibitem{Stephens}Stephens, A.D. \& Schierle, G.S.K. (2019) The role of water in amyloid aggregation kinetics. Curr. Opin. Struct., 58, 115-123.
\bibitem{Eisenberg}Eisenberg, D. \& Jucker, M. (2012) The amyloid state of proteins in human diseases. Cell, 148(6), 1188-1203.
\bibitem{Verma}Verma, P.K., Rakshit, S., Mitra, R.K. \& Pal, S.K. (2011) Role of hydration on the functionality of a proteolytic enzyme $\alpha$-chymotrypsin under crowded environment. Biochimie, 93(9), 1424-1433.
\bibitem{Jamal}Jamal, S., Kumari, A., Singh, A., Goyal, S. \& Grover, A. (2017) Conformational ensembles of $\alpha$-synuclein derived peptide with different osmolytes from temperature replica exchange sampling. Front. Neurosci., 11, 684.
\bibitem{Uversky}Uversky, V.N., Gillespie, J.R. \& Fink, A. L. (2000) Why are ``natively unfolded” proteins unstructured under physiologic conditions?. Proteins Struct. Funct. Genet., 41(3), 415-427.
\bibitem{Matthaeus}Matthaeus, F. (2009) The spread of prion diseases in the brain—models of reaction and transport on networks. J. Biol. Syst., 17(04), 623-641.
\bibitem{Liu}Liu, R.N. \& Kang, Y.M. (2020) Stochastic master equation for early protein aggregation in the transthyretin amyloid disease. Scientific Reports, 10(1), 12437.
\bibitem{Gillespie}Gillespie, D.T. (1977) Exact stochastic simulation of coupled chemical reactions. J. Phys. Chem. 81(25), 2340-2361.
\bibitem{McQuarrie}McQuarrie, D.A. (1967) Stochastic approach to chemical kinetics. J. Appl. Probab., 4(3), 413-478.
\bibitem{Gardiner}Gardiner, C.W. (1985). Handbook of stochastic methods. Vol. 3. Berlin: Springer.
\bibitem{Kolmogorov}Kolmogorov, A.N. (1938) On analytic methods in probability theory. Uspekhi Mat. Nauk, 5, 5-41.
\bibitem{Kampen}Van Kampen, N.G. (1992) Stochastic processes in physics and chemistry (Vol. 1). Elsevier.
\bibitem{Singh}Singh, M.S., Singh, M.K. \& Singh, R.B. (2024). Stochastic approach to study the properties of the complex patterns observed in cytokine and T cell interaction process. Nonlinear Dynamics, 112(3), 2237-2252.
\bibitem{Fano}Fano, U. (1947) Ionization yield of radiations. II. The fluctuations of the number of ions. Phys. Rev., 72(1), 26.
\bibitem{Rajdl}Rajdl, K., Lansky, P. \& Kostal, L. (2020) Fano factor: a potentially useful information. Front. comput. neurosci., 14, 569049.
\bibitem{Chanu}Chanu, A.L., Bhadana, J., \& Singh, R.B. (2020) Stochastic fluctuations as a driving force to dissipative non-equilibrium states. J. Phys. A Math. Theor., 53(42), 425002.
\bibitem{Bratsun}Bratsun, D., Volfson, D., Tsimring, L.S. \& Hasty, J. (2005) Delay-induced stochastic oscillations in gene regulation. Proc. Natl. Acad. Sci. 102 14593-14598.
\bibitem{Barrio}Barrio, M., Burrage, K., Leier, A. \& Tian, T. (2006) Oscillatory Regulation of Hes1: Discrete Stochastic Delay Modelling and Simulation. PLoS Comput. Biol. 2(9): e117. DOI: 10.1371/journal.pcbi.0020117.
\bibitem{Cai}Cai, X. (2007) Exact stochastic simulation of coupled chemical reactions with delays. J. Chem. Phys. 126 124108-1.
\bibitem{Anderson}Anderson, D.F. (2007) A modified next reaction method for simulating chemical systems with time dependent propensities and delays. J. Chem. Phys. 127.21.

\end{thebibliography}
\end{document}